\def\eg{e.g.\ }

\def\half{\frac{1}{2}}
\def\c2{\chi ^2}

\documentstyle[12pt,psfig,aaspp]{article}

\begin{document}

\title{A Statistical Investigation of the Contamination of 
Binary Gravitational Lens Candidates by Cataclysmic Variables} 

\author{Ann B.\ Saust}
\affil{Princeton University Observatory \\ Princeton, NJ 08544-1001, USA.}

\begin{abstract}
A common remark at conferences and meetings has been: ``One can fit {\em any}
light curve with binary lens models because of the large number of parameters!''
It is especially tempting to claim that some of the possible binary events in 
the Galactic microlensing searches could be caused by cataclysmic variables 
since both phenomena exhibits asymmetric light curves.
This paper presents a statistical analysis comparing the light curves of 
cataclysmic
variable stars to the light curves generated from binary lens models. It is 
shown that the light curves of high amplitude cataclysmic variables are
one example of
light curves that can {\em not} be explained with binary lens models. It is,
therefore, not only unlikely, but also impossible that cataclysmic variables
can contaminate the binary microlens candidates.
\end{abstract}



\section{Introduction}

Applications of gravitational lens theory was recently expanded when 
first Paczy\`{n}sky (1986) predicted the possibility of detecting dark
matter in the Galactic halo by searching for microlensing effects on
millions of stars in the LMC, and, by now, the four
collaborations: The MACHO group (for the latest results see the world 
wide web at: 
wwwmacho.mcmaster.ca/ or wwwmacho.anu.edu.au/), the EROS group 
(Aubourg et al.\ 1993), 
the OGLE group (see www.astro.princeton.edu/\~{ }ogle/ or www.astrouw.edu.pl), 
and the DUO group (Alard 1995) have detected more than 100
microlensing candidates. To distinguish
between variable stars and microlensing a number of criteria was used, one
of which was that the light curve be symmetric, i.e.\ only microlensing
by a single object was considered.

However, as the observations have shown, binary lenses do indeed exist; 
notable examples being the DUO 2 (Alard et al.\ 1995), OGLE 7 (Udalski
et al.\ 1994), and the LMC binary event (Bennett et al.\ 1996) candidates. 
A priori, binary lensing is expected in about 10\% of lensing events 
(Mao \& Paczy\'{n}ski 1991), and Mao \& Di Stefano (1995) developed a method 
that determines the lensing parameters from a given light curve. 
This method was successfully applied to DUO 2 (Alard et al.\ 1995) and OGLE 7
(Udalski et al.\ 1994) showing that their light curves are fit well by 
binary lens models; 
thereby, raising the question of how many of the objects previously believed to 
be variable stars could, in fact, be microlensing candidates. In addition,
Dominik \& Hirshfeld (1995) showed that the light curve of the MACHO LMC \# 1 
candidate could be explained better by a binary lens model than a single
lens model. A recent review of the Galactic microlensing searches can
be found in Paczy\'{n}ski (1996).

In an investigation similar to this one, Della Valle \& Livio (1995) found that 
the distribution of the amplitudes relative to
the durations of ordinary dwarf nova eruptions is significantly different
from those exhibited by the main body of the microlensing candidates from
the MACHO and OGLE collaborations. They concentrated on single lens models
which limits the investigation to cataclysmic variables with symmetric light curves.
They concluded that cataclysmic variables with symmetric light curves are not
a significant source of contamination, but speculated that the most serious
source of potential contamination may be due to old novae that exhibit dwarf
nova outbursts. By adopting the current estimates for the space density  of 
such old novae, they found that a significant fraction of the microlensing 
candidates could be cataclysmic variables.

In this paper, the possibility of confusing microlens candidates and
cataclysmic variables is investigated further, and,
since Della Valle \& Livio (1995)  estimated that only 20-40\% of cataclysmic 
variables have symmetric light
curves, only binary lens models, which can produce
asymmetric light curves, are considered.
Because of the well sampled light curve and its frequency-amplitude
relation, SS Cygni is used as a typical
cataclysmic variable, and the features of SS Cygni is compared to light curves
caused by binary microlenses.

\section{Basic Theory}

Since the theory of gravitational microlensing has been presented in detail
by numerous authors (e.g.\ Schneider et al.\ 1992 or Paczy\'{n}ski 1996), 
only the most basic ideas and notation are included in this section. 
Assume that a point source is located at
an angular diameter distance, $D_s$, and a binary lens at distance $D_l$
from the observer.
Let a light ray from the source pass at a distance $R_i$ from the $i'th$
lens with mass $M_i$. Define the dimensionless mass of the $i'th$ lens as:
\begin{equation}
m_i ~=~ \frac{(D_s ~-~ D_d)}{D_s D_d} \frac{4GM_i}{c^2}
\end{equation}
and define $m_1 ~+~ m_2 ~=~1$ which makes all angles measured in
units of the Einstein ring radius, $R_E$, which is given by:
\begin{equation}
R_E ~=~ \left ( \frac{4GD_{l}D_{ls}}{D_s c^2} \right ) ^{\half}
\end{equation}
for a lens with unit total mass. $D_{ls}$ is the angular diameter distance
between lens and source.
$R_E$ is a few AU for lensing in the LMC or the Galactic halo.
The lens equations for a binary lens then becomes:
\begin{eqnarray}
x_s & = & x ~-~ \frac{m_1 (x ~-~ x_1)}{r_1^2} ~-~ \frac{m_2 (x~-~x_2)}{r_2^2}
\label{lens1} \\
y_s & = & y ~-~ \frac{m_1 (y ~-~ y_1)}{r_1^2} ~-~ \frac{m_2 (y~-~y_2)}{r_2^2}
\label{lens2}
\end{eqnarray}
where $(x_s,y_s)$, $(x,y)$, and $(x_i,y_i)$ are the angles of
the source, the image
of the source, and $i'th$ lens, respectively, as seen by the observer, 
and $r_i$ is the impact
parameter in the observer plane. The theoretical light curves can then be
computed from the lens equations (\ref{lens1}) and (\ref{lens2}) using the
method described in Witt (1990) which simplifies the lens equations into
a fifth-order polynomial, which in turn can be solved numerically by Laguerre's
method (see Mao \& Di Stefano 1995).

\section{Numerical Method}

\begin{figure}
\centerline{\hspace{-0.4in}\psfig{figure=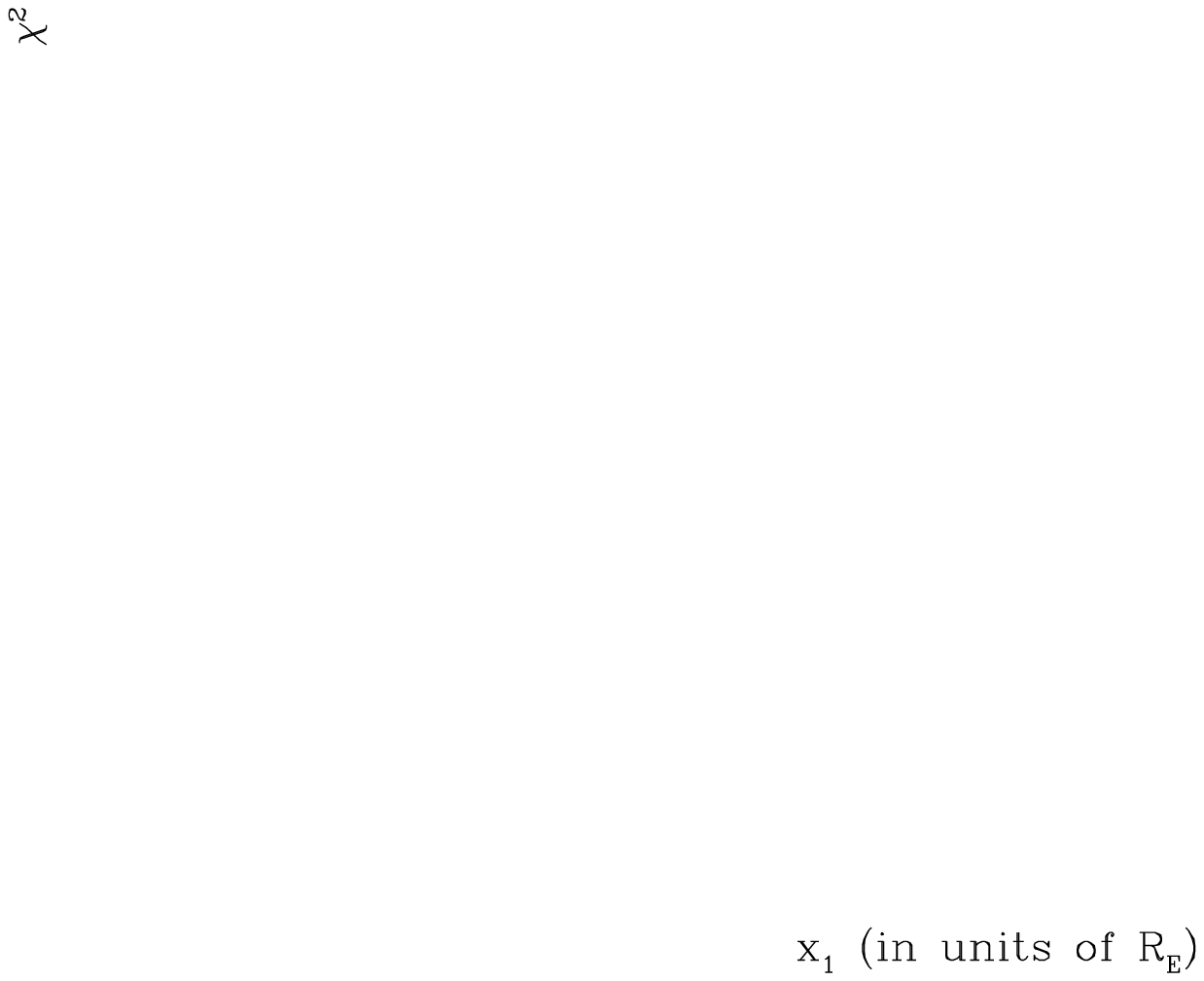,height=3.0in}}
\vspace{-3.2in}
\centerline{\psfig{figure=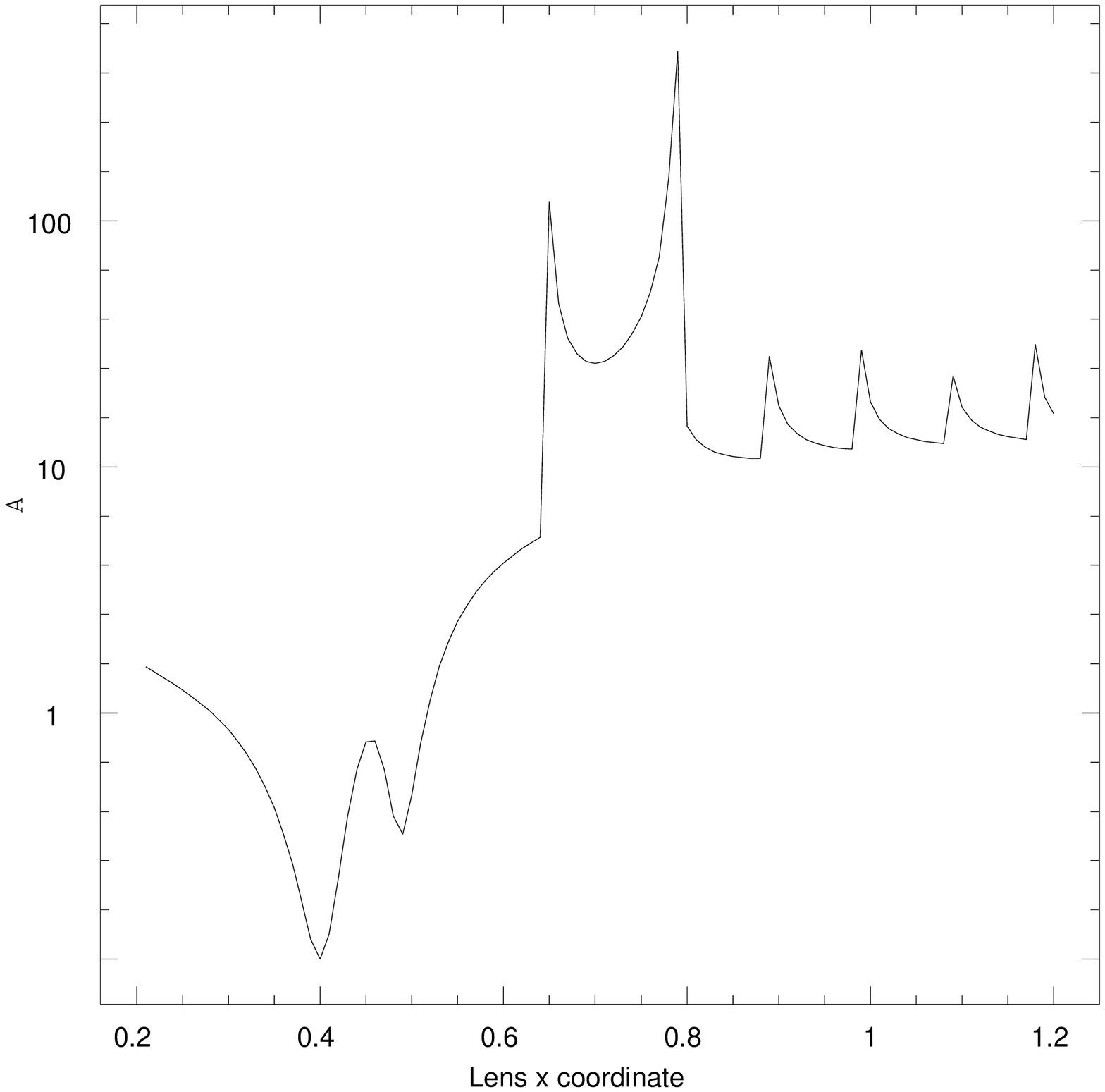,height=3.0in,bbllx=45bp,bblly=177bp,bburx=592bp,bbury=718bp,clip=t}}
\caption{$\c2$ as
a function of the x coordinate of one of the two lenses. Notice the many
local minima at $x_1$ = 0.51, 0.67, 0.87 etc.\ and the global minimum at 
$x_1$ = 0.40.
\label{param} }
\end{figure}

As can be seen from the previous section, the light curve of a background 
point source created by a binary gravitational lens
can be characterized by six variables: The (four) positions of the 
two lenses in the observer plane (($x_1$,$y_1$,) and ($x_2$,$y_2$) all measured
in units of $R_E$), the mass 
ratio of the lenses ($q ~=~ m_1 / m_2$), and the time parameter 
$T_E ~=~ R_E /V$ measured in days.
It is not necessary to solve for the source position, since changing ($x_s,~y_s$)
is equivalent to move the lenses. Therefore, define $y_s$ as 0.0 and
let $x_s$ vary between [-2.5,2.5]. In addition, it is necessary to know
the baseline, i.e.\ the magnitude of the unlensed source, and which
fraction of the observed light is caused by nearby unlensed sources. These
two parameters can be estimated by other means than including them as
free parameters, and they are, therefore, treated as input parameters. 
It is not
possible to analytically solve for the six parameters for a given light curve,
and it is, therefore, necessary to fit the observed and modeled light
curves numerically.
This can be done by minimizing $\c2$ given by
\begin{equation}
\c2 ~=~ \sum _{i=1}^{N_{obs}} \left ( \frac{\mu _i ~-~ \mu _{i,obs}}{\sigma _i} 
            \right ) ^2 ,
\end{equation} 
where $N_{obs}$ is the number of observations, $\mu _i$ and $\mu _{i,obs}$ are
the theoretical and observed {\em magnifications}, and 
$\sigma_i$ is the measurement
error. For large numbers of observations, $\c2$ is a normal distribution
with mean $N_{obs} ~-~ N_{parameters}$ and standard deviation 
$(2(N_{obs} ~-~ N_{parameters}))^{\half}$ (see e.g.\ Press et al., 1992).
Due to the nature of microlensing light curves, which can have several
sharp peaks appearing and disappearing with only small changes in the 
parameters, $\c2$ has many local minima. Using a simulated light curve (with $\sigma _i ~=~ 1$
for all $i$),
fig. \ref{param} shows how $\c2$
changes as a function of one parameter, the coordinate of one of the lenses.
As can be seen from the figure, the curve
shows numerous local minima which makes it impossible to use
standard numerical methods for minimizing a function (for instance the
Downhill-Simplex method or Simulated Annealing, which is even designed to 
disregard local minima, but fails in this case, see \eg Press et al.\ 1992).

Looking closely at fig.\ \ref{param} notice that: As the
function gets closer to the global minimum, the local minima get smaller. 
It would seems that a ``simulated tunnel
effect'' could jump from one local minima to another, since most of them are 
closely spaced.
Using these observations, a numerical method was developed that first
uses the Downhill-Simplex Method to find a (local) minima,
and then simulates a tunnel effect by randomly searching a six dimensional 
cube for nearby (local) minima
that are smaller than the present minima. To overcome features like the double
peak at $x_1 ~=~0.63$ and 0.79 in fig.\ \ref{param}, the cube is gradually 
expanded as long as no smaller minima is found. When a new (and smaller)
minima is found, the whole process repeats itself. The procedure is terminated
when the cube reaches a specified maximum size. 
When the cube is expanded
sufficiently slow and far, this method reliably finds the global minimum using
simulated light curves.
However, it turns out to be time consuming if the
initial guess of the six parameters is far from the global minimum.
To increase the efficiency of the program, the initial guesses are
determined by defining a grid in the 6 dimensional space, finding
the (local) minimum for each initial grid point, and then use the parameters
of the best minimum as initial guesses (which, depending on how fine the grid
is, may well be the global minimum).

\begin{figure}
\centerline{\psfig{figure=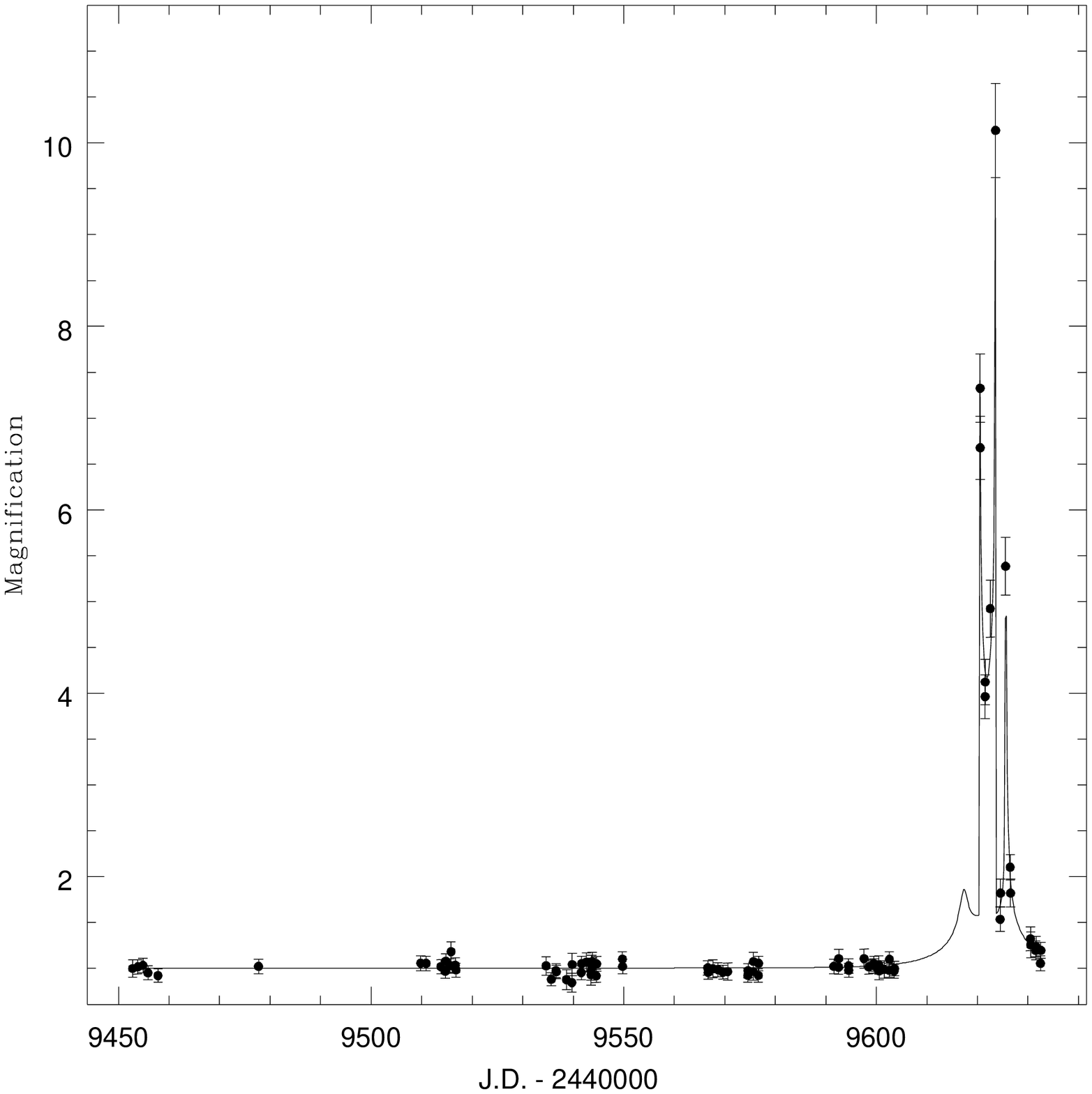,height=3.0in}\hspace{0.2in}
            \psfig{figure=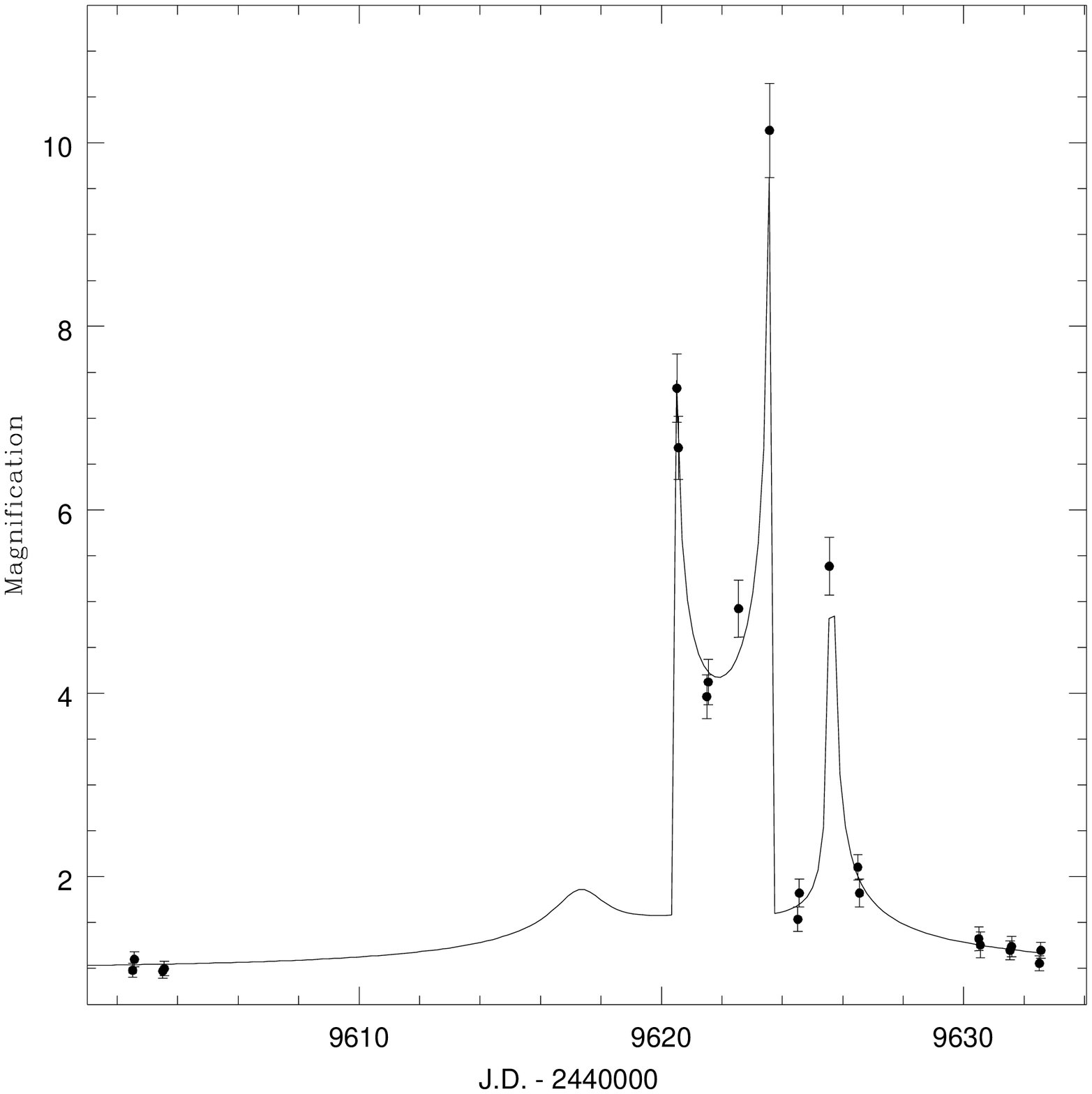,height=3.0in} }
\vspace{0.2in}
\centerline{(a)\hspace{3.2in}(b)}
\caption{The B band light curve of DUO 2. The solid line represents
the best model. (a) All the available data. (b) The lensed part of the
light curve.
\label{duo2} }
\end{figure}

As a test of the program, it was applied to the observations of DUO 2 
(Alard et al.\ 1995) in the B band, which has the most complex light curve
of the microlensing candidates so far, and, therefore, should be the most
difficult to fit. 
Assume a 30\% blend from an
unlensed source and adopt a baseline of $m_B ~=~ 21.45$ mag from the 
unlensed part of the light curve. The first attempts at fitting the
observations showed that the quoted measurements errors were overestimated,
and the errors were instead obtained from the scatter in the unlensed
part of the curve. In this case, the best fit has a 
$\c2$ of 78 for 80 data points (74 degrees
of freedom).
Using the same notation as above, the best fit parameters are:
\begin{equation}
\begin{tabular}{ll}
$x_1 ~=~ 0.42$ & $y_1 ~=~ 0.49$ \\
$x_2 ~=~ 0.54$ & $y_2 ~=~ -0.73$ \nonumber \\
$q ~=~ 0.33$ & $T_E ~=~ 8.85 {\rm ~days}$ \\
\end{tabular}  \nonumber
\end{equation}
from which the time of closest approach, $T_c$, can be calculated:
$T_c ~=~ 9621.7084$ J.D.\ - 2440000 or with the notation from Alard et al.\
(1995): $T_c ~=~ 85.06$ J.D.\ - 2449536.64861. For direct comparison, the
parameters used by Alard et al.\ (1995) are introduced: 
the distance between the two lenses, $a$ (in units
of $R_E$), the closest approach to the center of mass, $b$ (in units of $R_E$),
and the angle between the binary and the trajectory, $\theta$, this model yields:
\begin{equation}
a ~=~ 1.23 ~~~ b ~=~ 0.47 ~~~\theta ~=~ 92.9^{\circ}
\end{equation}

The observations and the best fit model are shown in fig.\ \ref{duo2}. 
This result is remarkably close to the solution found by Alard et al.\
(1995). The best fit parameters only deviate a few percent from each
other which is surprising since there's neither no guarantee that the global
minimum is found nor is the solution unique. In addition, Alard et al.\ (1995) 
fitted both the R and B band data, while only the B band data was considered 
here and included a finite source size.
It should also be noted that Alard et al.\ (1995) used a different
minimizing method, namely Levenberg-Marquardt's method. The Downhill-Simplex 
method was used here because it is more robust (but slower).
Notice that the method, described in this
paper, does not rely on any human decisions once the fitting parameters
(maximum size of cube etc.) are determined; the same program can be
applied to any light curve without changes or decisions on
the best guesses of the parameters. In particular, Alard et al.\ (1995) had to 
estimate $T_c$, and $T_E$ as an intermediate step. By fitting all six 
parameters simultaneously, this was avoided here.

\section{Observations}

Assume an unusual (and unlikely) scenario: Let the cataclysmic variable exhibit
one event during the observing run, let that run be shorter than normal for
microlensing searches, say one month, and assume that subsequent runs happens to
be in between bursts. This will give a worst case scenario for contamination of
binary lens candidates. Therefore, only cataclysmic variables 
with an outburst
frequency larger than one month are considered in the following.
The relation between amplitude and frequency is given by the Kukarkin-Parengo 
relation:
\begin{equation}
A _{min} - A _{max} ~=~ 0.70 \, [\pm 0.43] + (1.90 \, [\pm 0.22]) \, 
{\rm log}(T) \label{kuk}
\end{equation}
where $A _{min} - A _{max}$ is the amplitude of the burst in magnitudes
and $T$ is the elapsed time between bursts in days (see e.g.\ Warner 1995).
A one month frequency then corresponds to an amplitude of about 3.5 magnitudes
which again corresponds to a magnification of a factor of 25. Notice that,
according to eq.\ (\ref{kuk}), a low amplitude
variable would exhibit several events during the same observing run, and 
would, therefore, immediately be classified as a variable star.

\begin{figure}
\centerline{\psfig{figure=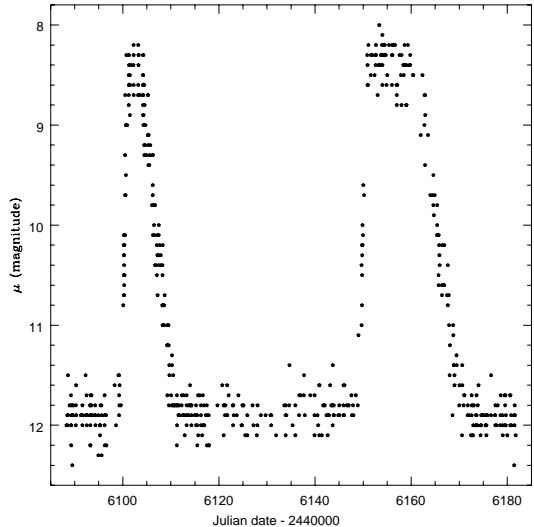,height=3.0in}}
\caption{A small part of the light curve of SS Cygni from 
Mattei et al.\ (1991).
\label{sscyg} }
\end{figure}

Based on the above, a good example of a cataclysmic variable is then
SS  Cygni, a dwarf nova of U Geminorum type, with an average outburst
frequency of 50 days and an amplitude corresponding to a magnification
of a factor of 30.
The American Association of Variable Star Observers combined observations
from numerous people (see Mattei et al.\ 1991, which can be found
at www.aavso.org/monographs.html, for details and
acknowledgments) thereby creating
a well sampled light curve of SS Cygni over a long period of time. A small
part of this light curve is shown in fig.\ \ref{sscyg}.

\section{Results}

\begin{figure}
\centerline{\psfig{figure=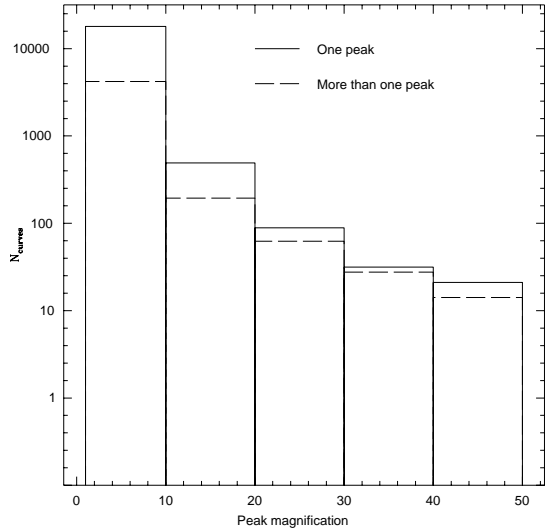,height=3.0in}}
\caption{The number of light curves, $N_{curves}$, as a function of the peak
magnification. Both light curves with one peak and two or
more peaks were included in this figure. 
\label{stat} }
\end{figure}

As we have seen in the previous sections, only high magnification events
(defined here conservatively as more than a factor of 10) are of interest. According to 
Schneider et al.\ (1992),
the probability that the magnification of a point source is larger than a 
certain value, $\mu _o$, is proportional to $\mu ^{-2}$ independent of 
lens model. Since only single peak light curves have any interest,
$\approx$ 250,000 light curves and their peak magnification using binary
lens models were calculated. The
resulting histogram is shown in fig.\ \ref{stat}. As can be seen from the
figure, of the total 250,000 light curves only $\approx$ 3\% are high
magnification events with one peak. The possibility of mistaking
cataclysmic variables as binary lens candidates is, therefore, exceedingly
small; out of the total of $\approx$100 microlensing candidates, one would
expect $\approx$10\% to show binary lens signatures (Mao \& Paczy\'{n}ski
1991), and only 3\% of these
would exhibit light curves similar to cataclysmic variables. 

\begin{figure}
\centerline{\psfig{figure=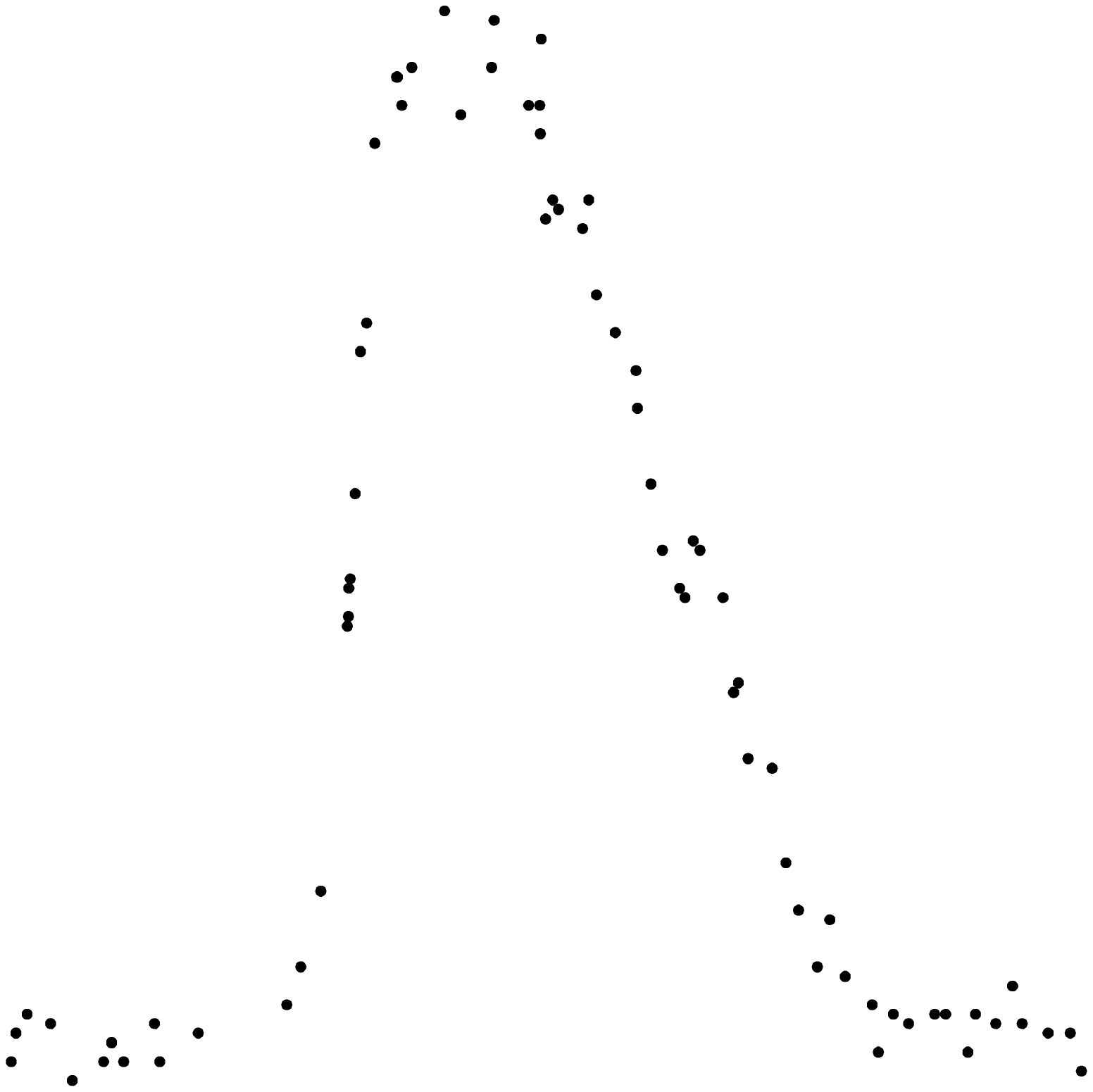,height=3.0in}\hspace{0.2in}
            \psfig{figure=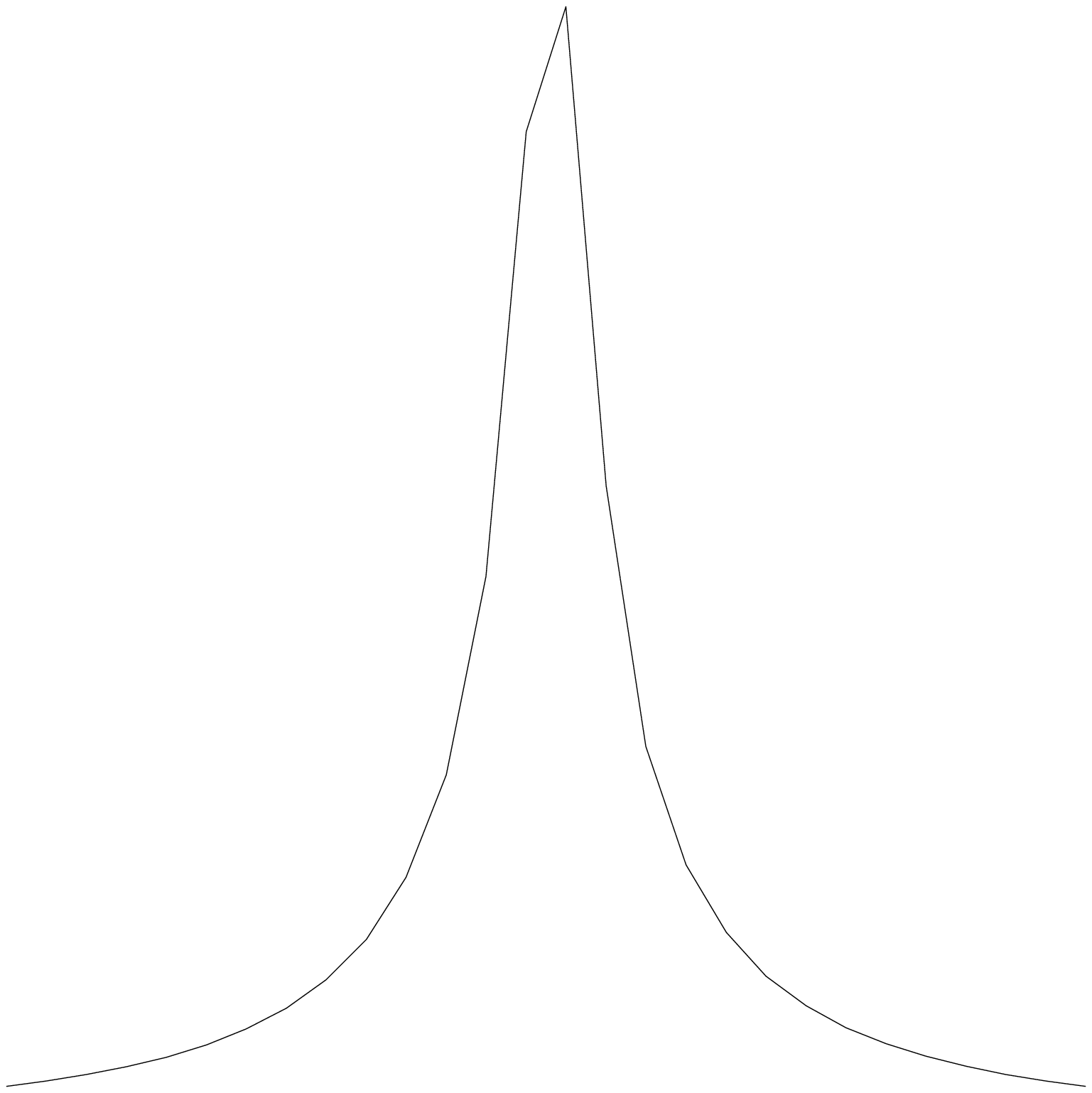,height=3.0in} }
\vspace{0.2in}
\centerline{(a)\hspace{3.2in}(b)}
\caption{The typical shape of an event for: (a) SS Cygni. (b) Models
with one peak and a peak magnification of 10 or more.
\label{shape} }
\end{figure}

\begin{figure}
\centerline{\psfig{figure=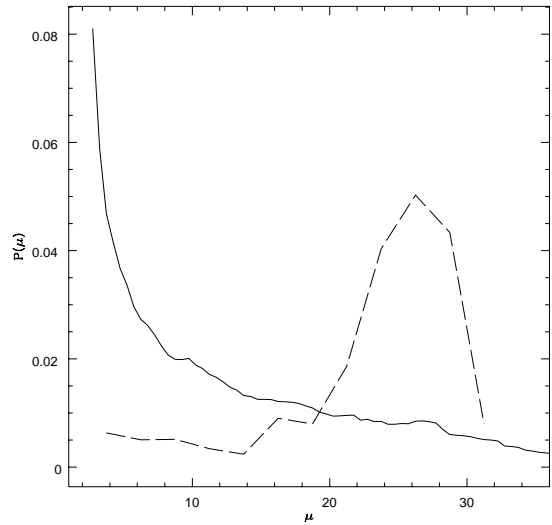,height=3.0in}}
\caption{The probability of the magnification being in the interval
[$\mu$ - 1.25, $\mu$ + 1.25] for models fitting the one peak, factor of ten
or more magnification criteria (solid line) and all the SS Cygni events 
from Mattei et al.\ (1991) (broken line).
\label{hist} }
\end{figure}
\begin{figure}
\centerline{\psfig{figure=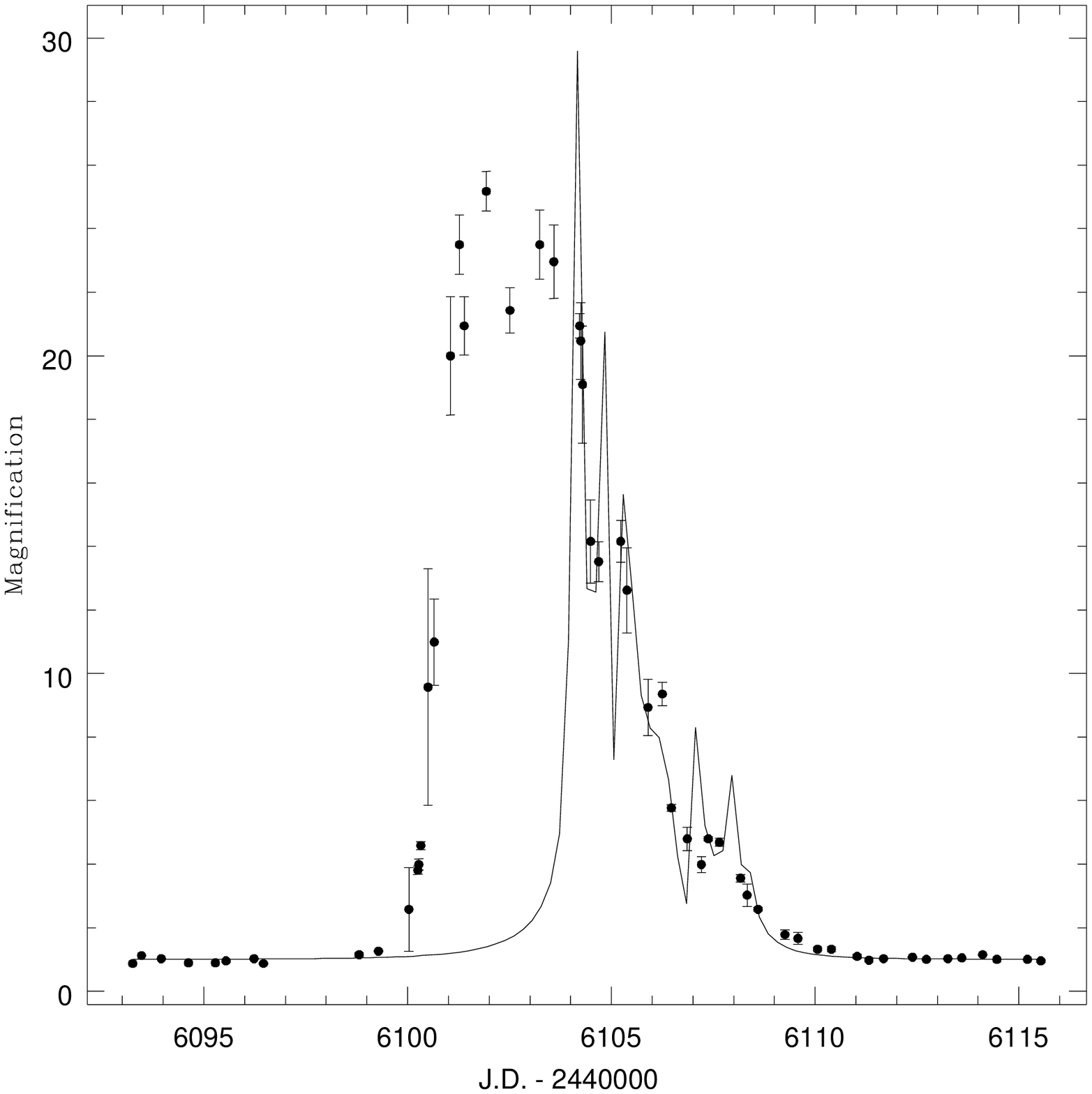,height=3.0in}}
\caption{The best fit to an event in SS Cygni.
\label{cygfit} }
\end{figure}

However, if
the light curve is well sampled, it is actually impossible to confuse the two
phenomena. Fig.\ \ref{shape} shows the shape of one event in SS Cygni
and a light curve of a high magnification event on the same scale. As
can be seen, the two light curves are incompatible with each other,
especially in the wings. To quantify this statement,
the probability of a magnification interval $\Delta \mu$, was calculated from
all events in Mattei et al.\ (1991) and all models fitting
the more than a factor ten, one peak criteria. To avoid numerical artifacts
from the large scatter in the SS Cygni observations, $\Delta \mu$ was set
equal to a magnification of a factor of 2.5 (corresponding to 1.0 magnitudes).
The resulting histogram is shown in fig.\ \ref{hist},
which clearly shows that the shapes of the light curves are 
incompatible with each other. 

As a final test, one event from SS Cygni's light curve was fitted
with a binary lens model. SS Cygni's light curve was smoothed by averaging 
over every four points thereby getting estimates of the measurement errors 
and reducing computation time by a factor of four. 
Not surprisingly, the ``best'' fit has a $\c2$ of 9,550 with 50 degrees of 
freedom, which clearly shows that the event in SS Cygni can not be modeled
by binary lenses. As can be seen from fig.\ \ref{cygfit},
the modeled light curve not only has more than one peak, 
but show no resemblance to the observations whatsoever. 
It is again concluded that cataclysmic variables can not be confused with
binary microlens candidates. 

\section{Conclusion}

It has been demonstrated that the developed numerical method reliably finds
the global minimum in all the simulated data. The program also determines
a best fit model to the binary lens candidate DUO 2 which is compatible with
the models published in Alard et al.\ (1995).

SS Cygni was chosen as an example of a typical cataclysmic variable star,
and it was demonstrated that not only is it highly unlikely that
the observed events can be explained by binary microlens models, but,
due to the difference between the shapes of the observed events and the
models, the two phenomena are incompatible with each other when the
light curve is well sampled. The attempt to fit one of the observed events
with a binary lens model was, as expected, a complete failure.

\acknowledgments {\em Acknowledgments:} I would like to thank B.
Paczy\'{n}ski for many helpful ideas, discussions, and comments on
this paper. I would also like to thank S. Mao for letting me use 
the DUO 2 observations,
for helpful discussions on the numerical method, and for supplying me with
a program that generates the light curves from eqs.\ (3) and (4). 
In this research, I have used and acknowledge with thanks data from the 
AAVSO International Database, based on observations submitted to the AAVSO 
by variable star observers worldwide. This work
was supported by NSF grant AST 93-13620.

\end{document}